\documentclass[12pt]{article}

\usepackage{amsmath}
\usepackage{graphicx}
\usepackage{geometry}

\usepackage{array}
\usepackage{booktabs}
\usepackage[normalem]{ulem}

\ifx\pdfoutput\@undefined\usepackage[usenames,dvips]{color}
\else\usepackage[usenames,dvipsnames]{color}
\IfFileExists{pdfcolmk.sty}{\usepackage{pdfcolmk}}{}
\fi
\usepackage[plainpages=false,pdfpagelabels,pagebackref=false,naturalnames=true,hyperindex=true]{hyperref}
\hypersetup{colorlinks=true,
urlcolor=Cerulean,linkcolor=BrickRed,citecolor=RoyalBlue,
  pdfpagemode=UseNone,
  pdfstartview=FitH}
\usepackage[all]{hypcap}

\usepackage{natbib}  

\title{Self-Organization and Artificial Life}
\author{Carlos Gershenson$^{1,2}$, Vito Trianni$^{3}$, Justin Werfel$^{4}$ \& Hiroki Sayama$^{5,6}$ \\
\mbox{}\\
{\scriptsize $^1$Universidad Nacional Aut\'{o}noma de M\'{e}xico, Mexico City, Mexico}\\
{\scriptsize $^2$ITMO University, St. Petersburg, Russian Federation}\\
{\scriptsize $^3$Institute of Cognitive Sciences and Technologies, Italian National Research Council, Rome, Italy}\\
{\scriptsize $^4$Wyss Institute for Biologically Inspired Engineering, Harvard University, Cambridge, MA 02138, USA}\\
{\scriptsize $^5$Center for Collective Dynamics of Complex Systems, Binghamton University, Binghamton, NY 13902, USA}\\
{\scriptsize $^6$Waseda Innovation Laboratory, Waseda University, Tokyo, Japan}\\
{\scriptsize cgg@unam.mx, vito.trianni@istc.cnr.it,
justin.werfel@wyss.harvard.edu, sayama@binghamton.edu}} 

\begin{document}
\maketitle

\begin{abstract}
{\color{black}Self-organization can be broadly defined as the ability of a system to display ordered spatio-temporal patterns solely as the result of the interactions among the system components. Processes of this kind characterize both living and artificial systems, making self-organization a concept that is at the basis of several disciplines, from physics to biology and engineering. Placed at the frontiers between disciplines, Artificial Life (ALife) has heavily borrowed concepts and tools from the study of self-organization, providing mechanistic interpretations of life-like phenomena as well as useful constructivist approaches to artificial system design. Despite its broad usage within ALife, the concept of self-organization has been often excessively stretched or misinterpreted, calling for a clarification that could help with tracing the borders between what can and cannot be considered self-organization. In this review, we discuss the fundamental aspects of self-organization and list the main usages within three primary ALife domains, namely ``soft'' (mathematical/computational modeling), ``hard'' (physical robots), and ``wet'' (chemical/biological systems) ALife. We also provide a classification to locate this research. Finally, we discuss the usefulness of self-organization and related concepts within ALife studies, point to perspectives and challenges for future research, and list open questions. We hope that this work will motivate discussions related to self-organization in ALife and related fields.}
\end{abstract}

\section{What is self-organization?}

{\color{black} The idea of self-organization can be traced to antiquity, including Greek and Buddhist philosophies~\citep{kirk1951natural,Gershenson2018}. The term ``self-organization'' was used sparingly in the 19th century, mainly applied to social systems.
Similar concepts had been proposed earlier by Kant~\citep{Juarrero1985}, and in the 1930s, it was introduced to embryology~\citep{Stengers1985}.}

The modern term ``self-organizing system'' was coined by \cite{Ashby1947sos} to describe phenomena where local interactions between independent elements lead to global behaviors or patterns.
The phrase is used when an external observer perceives a pattern in a system with many components, and this pattern is not imposed by a central authority among or external to those components, but rather arises from the collective behavior of the elements themselves. Natural examples are found in areas such as collective motion \citep{Vicsek2012}, as when birds or fish move in flocks or schools exhibiting complex group behavior; morphogenesis \citep{lawrence1992making}, in which cells in a living body divide and specialize to develop into a complex body plan; and pattern formation \citep{Cross1993} in a variety of physical, chemical, and biological systems \citep{CamazineEtAl2003,FeltzEtAl2006}, such as convection and crystal growth as well as the formation of patterns like stripes and spots on animal coats.

A formal definition of the term runs into difficulties in agreeing on what is a \emph{system}, what is \emph{organization}, and what is \emph{self} \citep{GershensonHeylighen2003a}, none of which are perfectly straightforward. However, a pragmatic approach focuses on when it is \emph{useful} to describe a system as self-organizing \citep{GershensonDCSOS}. This utility typically comes when an observer identifies a pattern at a higher scale but is also interested in phenomena at a lower scale; there then arise questions of how the lower scale produces the observables at the higher scale, as well as how the higher scale constrains and promotes observables at the lower scale. For example, bird behavior leads to flock formation, and descriptors at the level of the flock can also be used to understand regulation of individual bird behavior \citep{KeysDugatkin1990}.

Self-organization has been an important concept within a number of disciplines \citep{SkarCoveney2003}, including statistical mechanics \citep{Wolfram1983,Crutchfield2011}, supramolecular chemistry \citep{Lehn2017}, and computer science \citep{Kohonen2000,mamei2006case}.
Artificial Life (ALife) frequently draws heavily on self-organizing systems in different contexts \citep{Aguilar2014The-Past-Presen}, starting in the early days of the field with studies of systems like snowflake formation \citep{Packard1986} and agent flocking \citep{reynolds87flocks}, and continuing to the present day. However, there are often confusions and misinterpretations involved with this concept, possibly due to an apparent lack of recent systematic literature. 

In this work, we intend to:

\begin{enumerate}
\item Review research at the intersection of self-organization and ALife.
\item Provide a classification to locate this research.
\item Guide newcomers to the field with this classification.
\item Synthesize relevant concepts, challenges and open questions.
\item Open discussions on this topic within ALife and related fields.
\end{enumerate}

We first articulate some fundamental aspects of self-organization, outline ways the term has been used by researchers in the field, and then summarize work based on self-organization within \emph{soft} (simulated), \emph{hard} (robotic), and \emph{wet} (chemical and biochemical) domains of ALife. We then present a classification for categorizing different types of self-organization. We  also provide perspectives for further research. A list of open questions closes the paper.

\section{Usage}

Ashby coined the term ``self-organizing system'' to show that a machine could be strictly deterministic and yet exhibit a self-induced change of organization \citep{Ashby1947sos}. This notion was further developed within cybernetics \citep{vonFoerster1960,Ashby1962}. In many contexts, a thermodynamical perspective has been taken \citep{Haken1981,Haken1988,Jaffe2017}, where ``organization'' is viewed as a reduction of entropy in a (open) system \citep{NicolisPrigogine1977}. Since there is an equivalence between Boltzmann-Gibbs entropy and Shannon information, this notion has also been applied in contexts related to information theory \citep{Polani2003,Polani2008,Prokopenko:2008,Fernandez2013Information-Mea}. In this view, a self-organizing system is one whose dynamics lead it to decrease its information content, hence becoming more predictable. Based on information theory, the recent subfield of \emph{guided self-organization} explores mechanisms by which self-organization can be regulated for specific purposes --- that is, how to find or design dynamics for a system such that it will have particular attractors or outcomes \citep{Prokopenko:2009,Ay2012Guided-self-org,GSO2013,GSOInception2014,ProkopenkoGershenson2014}. For example, the self-organization of random Boolean networks \citep{Kauffman1969,Kauffman1993} can be guided to specific dynamical regimes \citep{Gershenson:2010}. 

There are several other definitions of self-organization as well. \cite{Shalizi2001} defines self-organization as an increase in statistical complexity, which in turn is defined as the amount of information required to minimally specify the state of the system's causal architecture. As an alternative to entropy, the use of the mean value of random variables has also been proposed \citep{Holzer:2011}. 

The concept of self-organization is also heavily used in organization science, with relevance to early artificial society models \citep{gilbert1995artificial,EpsteinAxtell1996} which have evolved into what is known today as computational social science \citep{lazer2009life}.

Self-organization is commonly used in a broad sense that encompasses self-assembly and other processes, but the term has at times been used in a more restrictive sense for far-from-equilibrium processes \citep{Moreno2009}. 



While there may be no single agreed-on definition of self-organization, this lack need not be an insurmountable obstacle for its study, any more than a lack of a unanimous formal definition of ``life'' has been an obstacle for progress in the fields of biology or ALife. In what follows, we provide a concise review of how self-organization has contributed to the progress of ALife.

\section{Domains}

One way to classify ALife research is to divide it into \emph{soft}, \emph{hard}, and \emph{wet} domains, roughly referring to computer simulations, physical robots, and chemical/biological research (including living technology as the application of ALife \citep{Bedau:2009}), respectively. Self-organization has played a central role in work in all three domains.

\subsection{Soft ALife}

\begin{figure}
    \centering
    \begin{tabular}{ll}
        A & B \\
        \includegraphics[width=0.3\columnwidth]{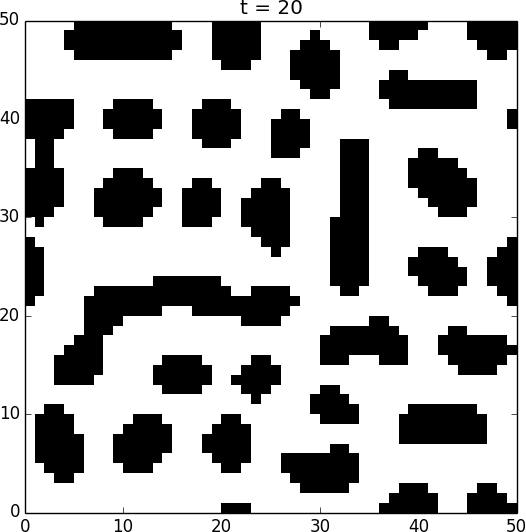} &
        \includegraphics[width=0.3\columnwidth]{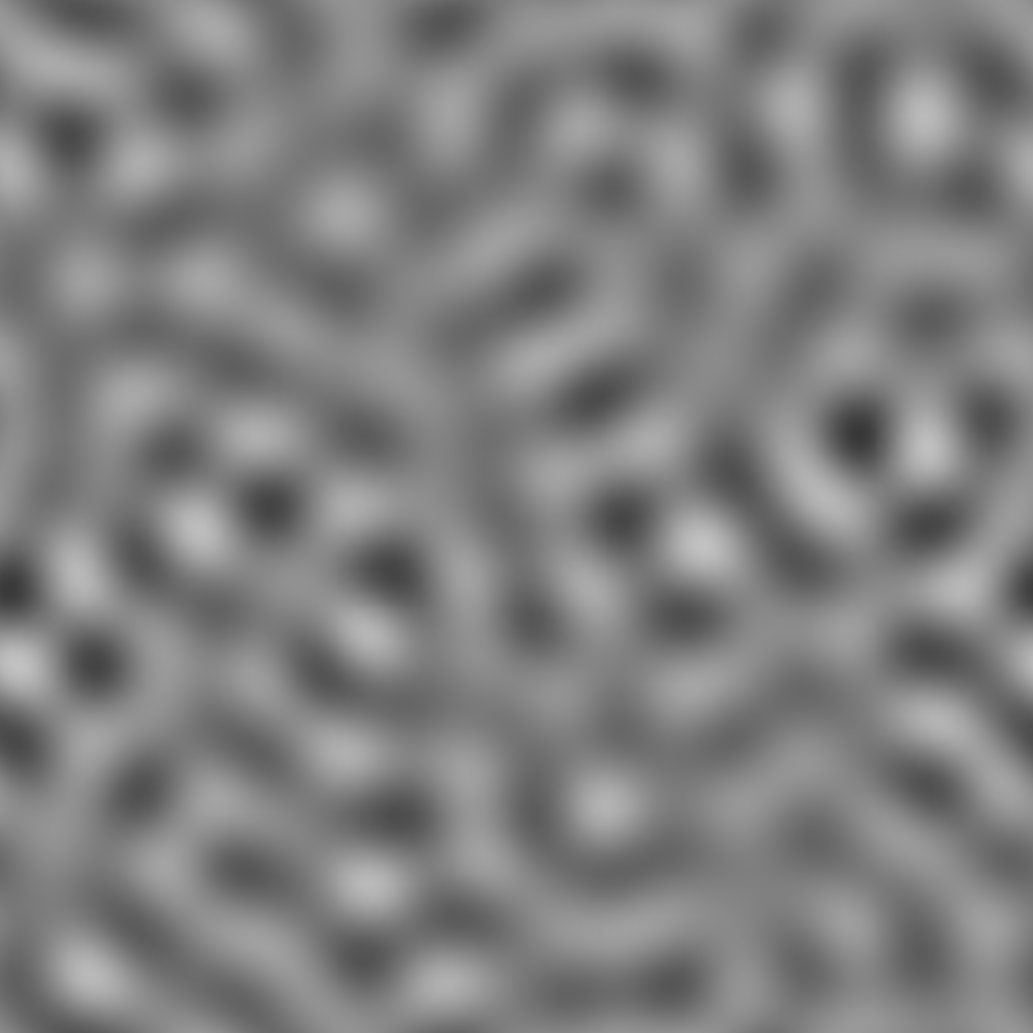}
    \end{tabular}
    \caption{Turing pattern formation \citep{Turing1952} as an illustrative example of self-organization in computational models. A: Simulation in CA using Young's discrete model \citep{Young1984}. B: Simulation in PDE using Turing's original formulation. Figures from \citep{Sayama2015}.}
    \label{fig:TuringCA-PDE}
\end{figure}

Soft ALife, or mathematical and computational modeling and simulation of life-like behaviors, has been linked to self-organization in many sub-domains.  Cellular automata (CAs) \citep{Ilachinski2001}, one of the most popular modeling frameworks used in earlier forms of soft ALife, are well-explored, illustrative examples of self-organizing systems. A CA consists of many units (cells), each of which can be in any of a number of discrete states, and each of which repeatedly determines its next state in a fully distributed manner, based on its current state and those of its neighbors. With no central controller involved, CAs can organize their state configurations to demonstrate various forms of self-organization: dynamical critical states such as in sand-pile models \citep{BakTangWiesenfeld1988} and in the Game of Life \citep{BakChenCreutz1989}, spontaneous formation of spatial patterns \citep{Young1984,Wolfram1984,ErmentroutEdelsteinKeshet1993} (Fig.\ \ref{fig:TuringCA-PDE}A), self-replication\footnote{Note that earlier literature on self-reproducing cellular automata \citep{vonNeumann1966,Codd1968} is not included here, because those models typically had a clear separation between a central universal controller and a structure that is procedurally constructed by the controller; thus they may not constitute a good example of self-organization as discussed in this article.} \citep{Langton1984,Langton1986,ReggiaEtAl1993,Sipper98}, and evolution by variation and natural selection \citep{Sayama1999,Sayama2004,SalzbergSayama2004,SuzukiIkegami2006,OrosNehaniv2007,OrosNehaniv2009}. Similarly, partial differential equations (PDEs), a continuous counterpart of CAs, have an even longer history of demonstrating self-organizing dynamics \citep{Turing1952,GlansdorffPrigogine1971,FieldNoyes1974,Pearson1993} (Fig.\ \ref{fig:TuringCA-PDE}B).

Another representative class of soft ALife that shows self-organization comprises models of collective behavior of self-propelled agents \citep{Vicsek2012}. Reynolds' ``boids'' model \citep{reynolds87flocks} is probably the best known in this category. In this work, self-propelled agents move in a continuous space according to three kinetic rules: cohesion (to maintain positional proximity), alignment (to maintain directional similarity), and separation (to avoid overcrowding and collision). A variety of related models have since been proposed and studied, including simplified, statistical-physics-oriented ones \citep{VicsekEtAl1995,LevineEtAl2000,AldanaEtAl2007,NewmanSayama2008} and more detailed, behavioral-ecology-oriented ones \citep{Couzin:2002,KunzHemelrijk2003,HildenbrandtEtAl2010}. These models produce natural-looking flocking/schooling/swarming collective behaviors out of simple decentralized behavioral rules, and they also exhibit phase transitions between distinct macroscopic states. They have also been used as inspiration for a variety of optimization algorithms \citep{DorigoStuetzle2004,Kennedy:1995,pham2006bees,Yang09,Krishnanand:2009}.

Such collective behavior models have been brought to \emph{artificial chemistry} studies as well \citep{Dittrich2001Artificial-Chem,BanzhafYamamoto2015}, such as \emph{swarm chemistry}, its variants, and other similar models \citep{Sayama2008Swarm-Chemistry,KreyssigDittrich2011,Sayama2011,Sayama2012,ErskineHerrmann2015,schmickl2016life,nishikawa2018exploration}, in which kinetically and chemically distinct species of idealized agents interact to form nontrivial spatiotemporal dynamic patterns. More recently, these collective behavior models have also been actively utilized in \emph{morphogenetic engineering} \citep{Doursat2011,DoursatEtAl2012}, in which researchers attempt to achieve a successful merger of self-organization and programmable architectural design, by discovering or designing agent rules that result in specific desired high-level patterns.

Other examples of self-organization in soft ALife are found in simulation models of artificial societies. Their roots can be traced back to the famous segregation models developed by Sakoda and Schelling back in the early 1970s \citep{Sakoda1971,Schelling1971,Hegselmann2017}, in which simple, independent decision making by individual agents would eventually cause a spatially segregated state of society at a macroscopic level. Agent-based simulation of artificial societies has been one of the core topics discussed in the ALife community \citep{epstein1996,Lansing2002}, and has elucidated self-organization of issues in social order such as geographical resource management \citep{LansingKremer1993,Bousquet2004313}, cooperative strategies \citep{LindgrenNordahl1993,Brede2011,AdamiEtAl2016,IchinoseSayama2017}, and common languages \citep{Steels1995,Kirby2002,SmithEtAl2003,LipowskaLipowski2012}. The literature on adaptive social network models may also be included in this category \citep{gross2009adaptiveNets,van2009emergence,BrydenEtAl2010,GeardBullock2010,davies2011if,sayama2020beyond}, as those ``artificial society'' models describe self-organization of society into a non-trivial configuration through co-evolution of autonomous dynamic state changes of social constituents and topological changes of social ties.

As adaptive networks at an individual organism level, brains and nervous systems also have been described as self-organizing systems for decades \citep{kelso1997dynamic,hesse2014self}, as neurons interact to produce behavioral and cognitive patterns. Self-organization of such neural systems has been particularly useful in computer science, and in the study of artificial neural networks \citep{gershenson2003artificial}; as a particularly conspicuous example, Kohonen networks \citep{Kohonen2000} are also called self-organizing maps. Since a large part of soft and hard ALife research deals with agents, animats or robots (virtual or physical) being controlled by artificial neural networks, it can be said that self-organization is present not only at the behavioral level, but also at the controller level in many cases.

Similar approaches have also been used in search and optimization techniques \citep{Downing2015Intelligence-Em}. For example, Watson and colleagues have proposed to use Hebbian learning \citep{hebb1949organization} to self-organize components of a complex system to resolve conflicts \citep{Watson:2010,Watson2011Global-Adaptati}. This mechanism probably has also been exploited beyond neural systems, as computational anthropology studies suggest \citep{Froese2014Teotihuacan,FROESE2018862}.

\subsection{Hard ALife}

Robots can be considered to be life-like artifacts in their ability to sense their physical environment and take action in response. Physical agents, even very simple ones, can evoke in the observer a particularly strong sense of being animate. From W. Grey Walter's tortoises \citep{walter1950,walter1951machine}, to simple machines based on the principles of Braitenberg's vehicles \citep{Braitenberg:1986}, from behaviour-based reactive robots \citep{Brooks1989} to recent biomimetic and bioinspired designs \citep{SaranliEtAl2001,WoodEtAl2013,KimWensing2017}, ALife built into machines stems from the rich dynamics underlying the interaction between the embodied agent and its environment, so that even simple mechanisms and behavioral rules can confer sophisticated life-like attributes to limited machines \citep{Simon1969}. Complex ALife forms can be attained either by increasing the sophistication of a single robot, or by increasing the number of robots in a system that, through the resulting interaction and self-organization, can then display more sophisticated abilities collectively, from adaptive responses to group decision making.

Hardware has the strong advantage that the physical characteristics of the system (dynamics, sensor performance, actuator noise profiles, etc.)\ are by definition realistic, whereas simulations are necessarily simplified and typically fail to capture phenomena that only become evident through material experimentation \citep{BrooksMataric1993,Jakobi1997,RubensteinEtAl2014}. Conversely, while simulation can readily handle very large numbers of agents, hardware considerations (cost, space, scalability of operation, etc.) have traditionally limited hard ALife studies to using a small number of robots. In some scenarios, self-organizing phenomena of interest do not necessarily require many robots. When the mechanism for coordination is based on \emph{stigmergy} (persistent information left in a shared environment), the important element is a large number of interactions between robots and environment, and even a single robot could suffice to generate complex patterns \citep{BeckersEtAl2000,WerfelEtAl2014}. More recently, hardware advances have made it possible to conduct physical experiments with robots in numbers exceeding a thousand \citep{RubensteinEtAl2014}.

Physical experiments have been used to explore self-organizing phenomena in a variety of areas. Aggregation of objects has been studied from a physics perspective \citep{GiomiEtAl2013} in ways inspired by behavior observed in living systems, such as cockroaches or bees \citep{Halloy2007,GarnierEtAl2008,KernbachEtAl2009} and using controllers designed through automatic methods like artificial evolution \citep{DorigoEtAl2004,FrancescaEtAl2014}. Another topic is collective navigation, in which groups of robots coordinate their overall direction of motion and collectively avoid obstacles \citep{BaldassarreEtAl2007,TrianniDorigo2006,TurgutEtAl2008}.  The coordination of flying robots has also been explored using self-organization \citep{Viragh2016,Vasarhelyi2018}. In other studies, collective decision-making processes are determined by positive feedback from recruitment and negative feedback from cross-inhibition  \citep{ReinaEtAl2018,ValentiniEtAl2015,ScheidlerEtAl2016,GarnierEtAl2009,GarnierEtAl2013,KernbachEtAl2009,FrancescaEtAl2014,ValentiniEtAl2017}.
Self-assembly \citep{Whitesides2002} is another form of self-organization largely studied in hard ALife with self-assembling or self-reconfiguring robots \citep{MurataEtAl1994,GriffithEtAl2005,ZykovEtAl2005,DorigoEtAl2006,YimEtAl2007,AmpatzisEtAl2009,RubensteinEtAl2014,Slavkoveaau9178}.


\subsection{Wet ALife}

Wet ALife, or physico-chemical synthesis of life-like behaviors, extensively utilizes self-organization as its core principle. A classic example is the spatial pattern formation in experimentally realized reaction-diffusion systems, such as the Belousov-Zhabotinsky reaction \citep{vanag2001pattern,adamatzky2008universal} and Gray-Scott-like self-replicating spots \citep{lee1994experimental,Froese2014}, where dynamic patterns self-organize entirely from spatially localized chemical reactions. Similar approaches can also be taken by using microscopic biological organisms (\emph{e.g.}, slime molds) as the media of self-organization \citep{garfinkel1987slime,hofer1995dictyostelium,maree2001amoeboids,adamatzky2008universal,adamatzky2015would}.

In research on the origins of life, molecular self-assembly plays the essential role in producing protocell structures and their metabolic dynamics \citep{rasmussen2003bridging,hanczyc2003experimental,rasmussen2004transitions,Protocells2008}. Chemical autopoiesis such as dynamic formation and maintenance of micelles and vesicles \citep{luisi1989self,bachmann1990self,bachmann1992autocatalytic,walde1994autopoietic} may also be included in this context.

More recently, dynamic behaviors of macroscopically visible chemical droplets, a.k.a. \emph{liquid robots} \citep{vcejkova2017droplets}, have become a focus of active study in ALife. In this line of research, interactions among chemical reactions, physical micro-fluid dynamics and possibly other not-yet-fully-understood microscopic mechanisms cause self-organization of spontaneous movements \citep{hanczyc2007fatty,cejkova2014dynamics} and complex morphology \citep{vcejkova2018multi} of those droplets. Moreover, droplet-based systems have also been used to demonstrate artificial evolution in experimental chemical systems \citep{parrilla2017adaptive}.

Recently, there have been a few studies on the collective behavior of protocells \citep[\emph{e.g.}][]{Qiao:2017} and droplets  \citep{vcejkova2017droplets}. The potential chemical interaction space is extremely vast, so it is difficult to explore with traditional techniques. Still, the automation of this exploration offers a promising approach \citep{GROMSKI20204}.

{\color{black} Wet ALife has developed more recently than the soft and hard perspectives, but it has a great potential to better understand living processes and also to exploit and regulate them with engineering principles and purposes.
}

\section{A Classification}

There are different potential classifications that could be considered to characterize self-organization in the context of ALife studies. 
One fundamental aspect concerns the level at which self-organization takes place with respect to the life-like process under consideration. In this respect, we can distinguish between \emph{internal} and \emph {external} self-organization. Internal self-organization would occur within an individual or agent, and could be functional to the production of life-like properties (\emph{e.g.}, morphogenesis) as well as useful to determine physical characteristics or behavioral responses that determine the way in which the individual agent interacts with its environment (\emph{e.g.}, pattern formation, neural plasticity). 
External self-organization is that occurring among individuals or agents. Such forms of self-organization pertain to the social aspects of life-like processes, which are often  fundamental to support reproduction and survival. These include collective behavior, social coordination, and ecological organization.
Note that in some cases the same process could be considered internal or external, depending on the observation level. For example, morphogenesis would be external at the cell level, but internal at the organism level. Behavior can be external at the individual level, but internal at the social level. 

An orthogonal direction that characterizes a self-organizing system concerns the nature of the interactions among the system components that bring about the life-like spatio-temporal patterns. In this respect, it is customary to distinguish between \emph{direct} and \emph{indirect} (\emph{e.g.}, stigmergic) forms of interaction.  When elements, individuals, or agents interact directly, their coordination can be fast. However, they need to be synchronized in time and space, and this sometimes can be challenging. Additionally, mechanisms must be concurrently provided for interactions to be encoded into a communication act and then suitably decoded. Indirect, stigmergic interactions take place by means of traces left in the environment, usually as a result of a unit of work performed by some agent that is recognized by a fellow agent \citep{TheraulazBonabeau1999}. Initially used to describe the organization of work in social insects (\emph{e.g.}, nest construction in termites or pheromone communication in ants), the concept of indirect interactions has been expanded to include any external medium that can store information and thus allow for coordination without the need of synchronous, direct communication. Indeed, the persistence of indirect interactions within the environment facilitates asynchronous coordination and the stratification of information, which can lead to complex patterns that extend in space and time. Note that internal self-organization is usually direct. This is because the environment in most cases is considered external to agents.

Examples of different types of self-organization belonging to different domains are included in Table~\ref{tab:examples}.

\begin{table}[!htbp]
\centering
\caption{Examples of ALife systems classified according to different types of self-organization and domains.  Depending on the observer purposes, the same ALife system could be considered as exhibiting different types of self-organization. Therefore, the types are non-exclusive and the boundaries between them are not sharp.}
\begin{tabular}{p{0.13\textwidth}|p{0.25\textwidth}|p{0.25\textwidth}|p{0.25\textwidth}}
\toprule
Level &  \multicolumn{1}{c}{Internal} & \multicolumn{2}{|c}{External}\\
\midrule
Interactions   & \multicolumn{2}{c|}{Direct}    & \multicolumn{1}{c}{Indirect}\\
\midrule
Soft   &  pattern formation, cellular automata , artificial neural networks & boids \citep{reynolds87flocks}, swarm chemistry \citep{Sayama2008Swarm-Chemistry}  & ant colony optimization \citep{DorigoStuetzle2004} \\
Hard   &  self-modeling robots  \citep{Bongard:2006}, swarm-bot \citep{DorigoEtAl2006} & Alice \citep{GarnierEtAl2008}, Jasmine \citep{KernbachEtAl2009}, Kilobots  \citep{RubensteinEtAl2014}  & TERMES \citep{WerfelEtAl2014} \\
Wet   &  protocells \citep{Protocells2008}, active droplets \citep{vcejkova2017droplets}  & xenobots \citep{Kriegman2020}, predator protocells \citep{Qiao:2017},  collective behavior of droplets \citep{vcejkova2017droplets}  & slime mold machines \citep{Adamatzky:2010} \\
\bottomrule
\end{tabular}
\label{tab:examples}
\end{table}


\section{Perspectives}
\label{sec:disc-concl}

As already mentioned above, we can understand a self-organizing system as one in which organization increases in time, without an external agency imposing this change. However, it can be shown that, depending on how the variables of a system are chosen, the same system can be said to be either organizing or disorganizing \citep{GershensonHeylighen2003a}. Moreover, in several examples of self-organization, it is not straightforward to identify the \emph{self} of the system, as oftentimes all elements composing the system can be ascribed equal agency. Finally, in cybernetics and systems theory, the dependency of the boundaries of a system on the observer has thoroughly been discussed \citep{Gershenson2013The-Past-Presen}: one wants to have an objective description of phenomena, but descriptions are necessarily made by observers, making them partially subjective.

It becomes clear, then, that discussing self-organization requires the identification of what is \emph{self} and what is \emph{other}, and what are the elements that are increasing in their \emph{organization}. Similar issues have been tackled by \cite{MaturanaVarela1980} in the definition of living systems as autopoietic systems. According to this tradition, a living system is inherently self-organizing because the \emph{self} is continuously produced or renewed by processes brought forth by the system's internal components. In other words, an autopoietic system can be recognized as a unity with boundaries that encompass a number of simpler/elementary components that are at the basis of the organization of the system, as they are responsible for the definition of the system boundaries and for the (re)production of the very same components \citep{varela1974autopoiesis}. This is a peculiar characteristic of living systems. If life is deeply rooted in self-organization, so can be ALife, and the several acceptations of ALife discussed above demonstrate the richness of the links it holds with self-organization. Nevertheless, autopoiesis did not originally consider evolution (history), an essential aspect of biology.

Whether evolution itself is an example of self-organization warrants discussion, too. Evolution is often depicted as synonymous with adaptation, a convergent process toward optimal types that are driven by external mechanisms (selection criteria or fitness landscapes). This has often been discussed as opposed or complementary to self-organization, most notably by \cite{Kauffman1993} and \cite{gould1990wonderful}. Meanwhile, there is also an effort of re-describing biological evolution as a kind of self-organization \citep{weber1996natural}, as all the mechanisms of evolution, such as variation, reproduction and selection, are ultimately grounded upon local, uncontrolled physical/chemical processes. Also, if one uses a very large spatial/temporal-scale perspective to observe evolution, it can be regarded as a self-organizing process of the population of evolving organisms as they may spontaneously generate more diverse species, more complex inter-specific interactions, and even higher-order evolving entities, as  diverse scales of space, time, and complexity are relevant \citep{levin2005self}.


Looking at the perspectives of ALife, it can be useful to think of self-organization as a common language that unifies the soft, hard and wet domains. {\color{black} The term is broadly used across many areas, pointing to the existence of common features that can tie together otherwise disparate studies. By recognizing and exploiting these commonalities, a better understanding of self-organization should help the advancement of ALife.}
The ALife community can progress owing to shared concepts and definitions, and despite the mentioned difficulties, self-organization stands as a common ground on which to build shared consensus. Most importantly, we believe that the identification and classifications of the \emph{mechanisms} that underpin self-organization can be extremely useful to synthesize novel forms of ALife and gain a better understanding of life itself.

These mechanisms should be identified at the level of the system components and characterized for the effects they have on the system organization. Mechanisms pertain to the modalities of interaction among system components (\emph{e.g.}, collisions, perceptions, direct communication, stigmergy), to behavioral patterns pertaining to individual components (\emph{e.g.}, exploration vs.\ exploitation), and to information enhancement or suppression (\emph{e.g.}, recruitment or inhibitory processes). The effects of the mechanisms should be visible in the creation of feedback loops --- positive or negative --- at the system level, which determine the complex dynamics underlying self-organization.
We believe that, by identifying and characterizing the mechanisms that support self-organization, the synthesis of artifacts with life-like properties would be much simplified. In this perspective, mechanisms underlying self-organization could potentially be thought of as \emph{design patterns} to generate ALife systems \citep{Babaoglu:2006hy,FernandezMarquez:2012es,Reina:2015hu}. By exploiting and composing them, different forms of ALife could be designed with a principled approach, owing to the understanding of the relationship between mechanisms and system organization.

The possibility of exploiting self-organization for design purposes is especially relevant toward the development of \emph{living technologies}, that is, technologies presenting features of living systems \citep{Bedau:2009}, such as robustness, adaptability, and self-organization, which can include self-reconfiguration, self-healing, self-management, self-assembly, etc., often named together as ``self-*'' in the context of autonomic computing \citep{Poslad2009}.

Self-organization has been used directly in living technologies within a variety of domains \citep{Bedau2013IntroductionLT}, from protocells \citep{Protocells2008} to cities \citep{Gershenson:2013}. Recent work programming \citep{Adamatzky:2010} or designing multicellular organisms \citep{doi:10.1063/1.5038337,Kriegman2020} also falls within this category.
Also, several methodologies that use self-organization have been proposed in engineering \citep{Frei:2011}. A major leap forward can be expected when principled design methodologies are laid down, and a better understanding of self-organization for ALife can be at the forefront of the development of such methods.

It is also worth considering when self-organization is \emph{not} useful in the context of ALife. Tracing a clear line across the domain is of course impossible, but our reasoning above provides some suggestions. Indeed, self-organization does not account for every life-like process, for instance when there is no clear increase in organization. For instance, hard ALife has strongly developed the concept of embodied cognition and morphological computation \citep{Pfeifer2007,Pfeifer:2009ba}, where the dynamics of mind-body-environment interaction are fundamental aspects. These dynamics, albeit very complex, are not easily described within the framework of self-organization.
{\color{black} Self-organization is useful when we are interested in observing phenomena at more than one scale, as it allows us to describe how elements interact to produce systemic properties. Still, if we are only interested in observing phenomena at a single scale, then perhaps self-organization would not offer any descriptive advantage. Examples include embodied cognition (when we are focusing on a single cognitive agent and its interaction with its environment) and most of the traditional types of evolutionary algorithms (when there are no interactions between individuals of a population).}

{\color{black} Depending on the desired function of a system and the properties of its environment, several balances have to be considered, e.g., between order and chaos, between robustness and adaptability, between production and destruction, between exploration and exploitation. Self-organization can be useful to let systems find by themselves the appropriate balances for their current context, as the optimal balance can change \citep{GershensonHelbing2015}.}

\section{Open Questions}

{\color{black} There are several open questions which make for promising lines of research in the near future within ALife:
\begin{enumerate}
\item \emph{How can self-organization be programmed?} Self-organization relies on interactions (direct or indirect). Thus, it makes sense to focus on designing interactions to regulate and guide self-organization. Mediators \citep{Michod2003,Heylighen:2006} can promote or constrain individual behaviors, precisely to achieve the proper interactions that will lead to the desired self-organization \citep{GershensonDCSOS}. Information-theoretical approaches can also be used to program self-organization \citep{GSOInception2014,Krakauer:2020}. Still, proposed approaches have been either too general or too specific. This makes it difficult to replicate successful self-organizing solutions beyond the original problems and remains an open challenge. 

\item \emph{Can the macroscopic outcomes of self-organization be predicted?} Interactions in complex systems generate novel information that is not present in initial nor boundary conditions, limiting predictability. This is referred to as ``computational irreducibility'' \citep{Wolfram:2002,Gershenson:2011e}: there is no shortcut for the future, a system has to go through all intermediate steps. Thus, \emph{a priori} claims are limited, and we often work with \emph{a posteriori} approaches. In some cases, coarse-grained descriptions can be found to predict self-organization and other properties \citep[\emph{e.g.},][]{PhysRevLett.92.074105}. Still, this has not been generalized. In the ALife community, we rely on the synthetic method \citep{Steels1993}: we build artificial systems to contrast theories, but of course this is \emph{a posteriori}. Even when prediction is limited by the complex nature of phenomena studied within ALife, forecasting could be useful. Just like with the weather, precise prediction is not possible (\emph{e.g.}, when, where, and how much will it rain?), but within a certain range, forecasts can be made with a high probability (\emph{e.g.}, $80\%$ chance of rain).

\item \emph{What is the role of self-organization in the open problems of ALife?} \cite{BedauEtAl2000} listed fourteen challenges grouped in three broad issues: the transition to life; the evolutionary potential of life; and the relation between life, mind, and culture. It can be argued that self-organization is present in all of these, and thus relevant. Certainly, solving questions related to self-organization will not solve all ALife questions, but it can provide useful advances. For example, research related to open-ended evolution \citep{taylor2016open} goes beyond self-organization. Still, better understanding self-organizing mechanisms could assist in the development and characterization of systems that exhibit open-ended evolution.

\item \emph{What are the theoretical and practical limits of self-organization?} Even when it has demonstrated its usefulness, self-organization is no panacea. Self-organization is most appropriate when there is multiscale causality and high complexity \citep{10.3389/frobt.2020.00041}, but centralized or distributed approaches can be more appropriate for other contexts (when there is only bottom-up or top-down causality, or when complexity is low or medium). Still, further work is needed to be able to identify qualitative and quantitative limits of self-organization.

\item \emph{How can understanding of self-organization in ALife benefit other disciplines?} These include biology, medicine, engineering, philosophy, sociology, economics, and more. Independent of whether ALife is credited or not, the question is whether ALife research will be able to contribute to the solution of problems that otherwise would not be solvable. There are promising examples and successful case studies \citep[\emph{e.g.}][]{Knight:2014,10.1371/journal.pone.0190100}, but broader adoption and dissemination are required to make a difference.

\end{enumerate}

}

These and more questions highlight the strong role that self-organization has within ALife. Searching for their answers will be challenging, but the insights provided will permeate beyond ALife.

\section*{Acknowledgements}

This article benefited from comments by Luis Rocha and reviewers from the ALIFE 2018 conference on an earlier version of this work \citep{GershensonALife2018}.

\footnotesize
\bibliographystyle{apalike}
\bibliography{refs,carlos,justin,hiroki-additional} 

\end{document}